\documentclass[twocolumn,trackchanges]{aastex63}
\usepackage{url}

\usepackage{longtable}
\usepackage{booktabs}
\usepackage{tabu}
\usepackage{graphicx}
\usepackage{multirow}
\usepackage{ulem}
\usepackage{color}
\usepackage{amsmath}

\received{X}
\revised{Y}
\accepted{Z}

\shorttitle{r-rpocess in CEJSN}
\shortauthors{Jin \& Soker}

\graphicspath{{./}{figures/}}

\begin{document}

\title{Robust r-process Nucleosynthesis Beyond Lanthanides in the Common Envelop Jet Supernovae}

\author{Shilun Jin}
\affiliation{Institute of Modern Physics, Chinese Academy of Sciences  \\
509 Nanchang Rd Lanzhou, Gansu, 730000, P. R. China}

\author{Noam Soker}
\affiliation{Department of Physics, Technion, Haifa, 3200003, Israel}

\date{\today}

\begin{abstract}
The common envelop jet supernovae (CEJSN) r-process scenario has been proposed as an r-process nucleosynthesis site in the past decade. Jets launched by a neutron star that spirals-in inside the core of a red supergiant star in a common envelope evolution supply the proper conditions for the formation of elements heavier than iron through the rapid neutron capture process.
The present work initially unveils the r-process abundance patterns that result from the density profile in the relatively long-lived jets. 
The results indicate that the CEJSN r-process scenario can produce the largest ratio of the third r-process peak elements to Lanthanides among current r-process scenarios, and in addition can form quite an amount of Lanthanides in a single event.  
The comparison of the ratio of the third peak elements to the Lanthanides with a number of observed r-enhanced metal-poor stars and with other r-process scenarios suggests that a high mass of third peak elements is anti-correlated with high fraction of Lanthanides, both in observations and theory.
The CEJSN r-process scenario plays a significant role in this conclusion, since it reproduces the observational features of some particular r-enhanced metal-poor stars where other r-process scenarios encounter problems.  
Due to the formation of extremely heavy elements, the CEJSN also offers a credible estimation on the age of the most Actinide boosted star by cosmochronometry.  

\end{abstract}

\section{Introduction} 
\label{sec:intro}

About half of the elements heavier than Iron are formed  in the rapid neutron capture process (r-process). However, the sites of this nucleosynthesis remain uncertain. After the discovery of the gravitational wave event GW170817(\cite{PhysRevLett-119-161101,Abbott_2017}) and its kilonova AT2017(\cite{coulter17,Villar_2017}), the binary neutron star (NS) merger (NSM) became the unique identified r-process site so far. The support comes from the Lanthanide bearing light curve(\cite{Chornock_2017}) and the inferred presence of other heavy unstable isotopes(\cite{kasen17}) that were found in later phase of this event(\cite{pian17,Sr19,kasliwal19}). The question of whether the NSM is the dominated or even the sole site of r-process nucleosynthesis is still controversial. The frequency of observed NSM events suggests that it barely explains the Lanthanide abundance in the solar system(\cite{Ji_2019,Waxmanetal2018,holmbeck_2023}). The NSM r-process by itself is insufficient to account for the observed Eu in the early Universe(\cite{Côté_2019}).
Other merger events like NS black hole (BH) mergers have been confirmed by gravitational wave detection, but electromagnetic counterparts from r-process are still undiscovered(\cite{Siegel22}). 

Core collapse supernovae (CCSN)(\cite{Burrow21}) have been a main candidate of r-process nucleosynthesis site since 1956(\cite{B2FH}).
The neutrino driven wind form the proto-NS in this site was believed to be a major source of heavy elements for many years.
Recent studies cast doubts on this scenario as a robust r-process site for the third r-process peak and the Actinide(\cite{Ebinger_2019}). It might only produce elements around the second r-process peak with the moderate neutron exposure that is usually thought to be a site of weak r-process(\cite{Truran2002,Montes2007}). 
Magnetohydrodynamic supernovas (MHDSN)(\cite{mosta2018,Yong2021,oberg2021_3D}), where the newly born NS in a CCSN launches jets, and collapsars(\cite{collapsar}), where the newly born BH in a CCSN launches jets, have been proposed as potentially major r-process sites for heavy elements since jets contain highly neutron-rich material in both types of objects.

Astronomical observations of old stars shed light on the main r-process sites in another aspect. These metal-poor stars(\cite{metalpoor}) are enriched by only one to a few r-process events and are termed as r-process enhanced stars(\cite{Holmbeck_2020}). They are classified into three sub-classes according to their r-process enhancement as follows. r-I: with $+0.3 < {\rm [Eu/Fe]} \leq +0.7$ and ${\rm[Ba/Eu]}<0.0$ \footnote{${\rm [A/B]} \equiv \log(N_{A}/N_{B})_{\rm star} -\log(N_{A}/N_{B})_{\rm sun}$, where $N$ is the number density of the element in the star or sun, respectively.}. r-II: with $+0.7 < {\rm [Eu/Fe]} \leq +2.0$ and ${\rm [Ba/Eu]} < 0.0$. r-III: with ${\rm [Eu/Fe]}> +2.0$ and [Ba/Eu] $<$ -0.5. 
Here r-I represents moderately r-enhanced stars, r-II and r-III are highly enhanced stars. The r-process isotopes in these r-enhanced stars that are synthesised in the main r-process site(s) reveal crucial features of the r-process nucleosynthesis. As all the models except NSM have no direct observational evidences, these features can significantly constraint and guide theoretical r-process models. 

In the past decade the common envelope jet supernova (CEJSN) r-process scenario has been  developed as a new potential r-process site(\cite{Papishetal2015, Grichener_2019}). 
At a late evolutionary phase of a massive binary system the initially more massive star has left a NS remnant after a CCSN explosion while the initially lower mass star turned to a red supergiant (RSG). 
The RSG expands and engulfs the NS that spirals-in inside the RSG envelop and then inside its core, i.e., a common envelope evolution (CEE). The NS launches jets as it accretes gas from the envelope and then from the core via an accretion disk. In the envelope the jets that the NS launches are too weak to support r-process nucleosynthesis. Once the NS enters the core of the RSG it accretes mass via an accretion disk at a very high rate and the energetic and dense jets can support r-process nucleosynthesis(\cite{Grichener_2019}). The core is destroyed by transferring a large fraction of its mass to form an accretion disk around the NS. The bipolar jets that carry a mass of $\simeq 0.1M_\odot$ and propagate at a speed of $\simeq 0.3-0.5c$ collide with the core material in a short period of time of $\approx 10-100~{\rm s}$ and eject the rest of the core material. 
Several studies on the formation, on the mechanisms and on the properties of CEJSN(\cite{Akashi_2021,Grichener_2022,10.1093/mnras/stab2233}) have demonstrated its potential as an r-process site.

In this work, the elemental abundance patterns of r-process in the CEJSN are initially presented. Both fairly considerable Lanthanide elements and very robust r-process third peak elements are synthesized synchronously. We firstly examine the ratio of elements in the third r-process peak to the Lanthanide on all available r-enhanced stars. The scattering distribution of this ratio might imply that the universal pattern of strong r-process no longer appears beyond elements heavier than Lanthanide, and multiple r-process sites are needed to account for the formation of elements in the third r-process peak and for the Actinides. By comparing the contribution to this ratio from other well known r-process sites, the CEJSN hence offers a significant contribution of r-process isotopes, and especially plays an irreplaceable role in the third peak region due to the density profiles in the propagating jets.

This letter is structured as fellows. In section \ref{sec:method}, we discuss our method for describing the density function of CEJSN and the network of nucleosynthesis evolution. Our results are described in Section \ref{sec:resul}, the abundance pattern and the remarkable feature, Lanthanides versus the elements in the third peak of r-process, are presented. Finally, the conclusions are given in the Section \ref{sec:concl}.

\section{METHODS}
\label{sec:method}

\subsection{Density Function}
\label{sec:den}

The density of the outflow from the CEJSN jets decreases exponentially with time $t$, but slightly different from the $t^{-3}$ power law in the neutrino-driven winds\cite{Arcones_2013} of CCSNe or the outflows encountered in NSM\cite{Radice_2018}. The $t^{-3}$ power law assumes a homologous expansion, namely, the expansion speed is proportional to distance $u=r/t$. 
In such a case the density in each parcel of gas decreases as $(t/\tau)^{-3}$ because the volume element of each parcel of gas that span a solid angle $\Omega$ is $dV=\Omega r^2 dr=\Omega t^3 u^2 du$, where we substituted $r=ut$.
In CEJSN jets the velocity behaves differently. In a long-lasting jet, that the timescale much larger than the local dynamical time scale, the velocity inside the jet is more or less constant, as the process is not explosive. 
In that case the distance between two points in a parcel of gas along the same radial direction is constant. Namely, the volume of the parcel of gas goes as $dV \propto t^2$. 
Therefore, the following density evolution with time to better represent the situation in CEJSN jets is shown in Eq.\ref{eq:DensityEX2}.
\begin{equation}  
\rho_{\rm cej}(t)= \rho_0 \times
\begin{cases}
\exp(-t/\tau), \qquad t \le \tau \\
\left( \frac{t}{\tau} \right)^{-2}, \qquad  t>\tau.
\end{cases}
\label{eq:DensityEX2}
\end{equation}

\subsection{Reaction Network}
\label{sec:net}

A large grid of nucleosynthesis calculations by sampling the parameter space of CEJSN jets has been performed with the nuclear reaction network code \texttt{SkyNet}\cite{skynet_2017} and employing the reaction rate database JINAREACLIB\cite{Cyburt_2010} consisting of 7836 nuclei. 
Three parameters of initial electronic fraction $Y_{e}$, initial entropy per baryon $s$, and expansion timescale $\tau$ are vaired in the calculations. 
The ranges of these parameters are consistent with the properties of the jets in the CEJSN scenario\cite{Grichener_2019}.
The entropy per baryon was varied in increments of 2$K_{B}$ from 2$K_{B}$ to 30$K_{B}$. The electron fraction was varied from  0.005 to 0.250 in equal steps of 0.01. The dynamical timescale was taken to be 2ms to 30 ms with step of 2ms. The initial temperature is 10 GK in the network evolution. 
For a given initial composition, initial entropy, and density as a function of time, \texttt{SkyNet} evolves the composition of the material and self-consistently incorporates changes in the entropy due to nuclear reactions.

\section{Results and Discussion}
\label{sec:resul}
\subsection{Abundance pattern}
\label{sec:abu}
The representative abundance patterns of CEJSN are selected in term of the universality(\cite{Frebel18})of r-process. 
It exhibits essentially consistent relative pattern for elements from Barium to Ytterbium, i.e., 56 $\le Z \le$ 70, in many r-process enhanced stars.
This remarkable feature of the r-process is so called universality, which is a solid proof of the r-process nucleosynthesis(\cite{Mumpower_2016}), although the exact mechanism is not fully understood.

\begin{figure}
\centering
\includegraphics[trim=0cm 0 0 0, clip, width=0.48\textwidth]{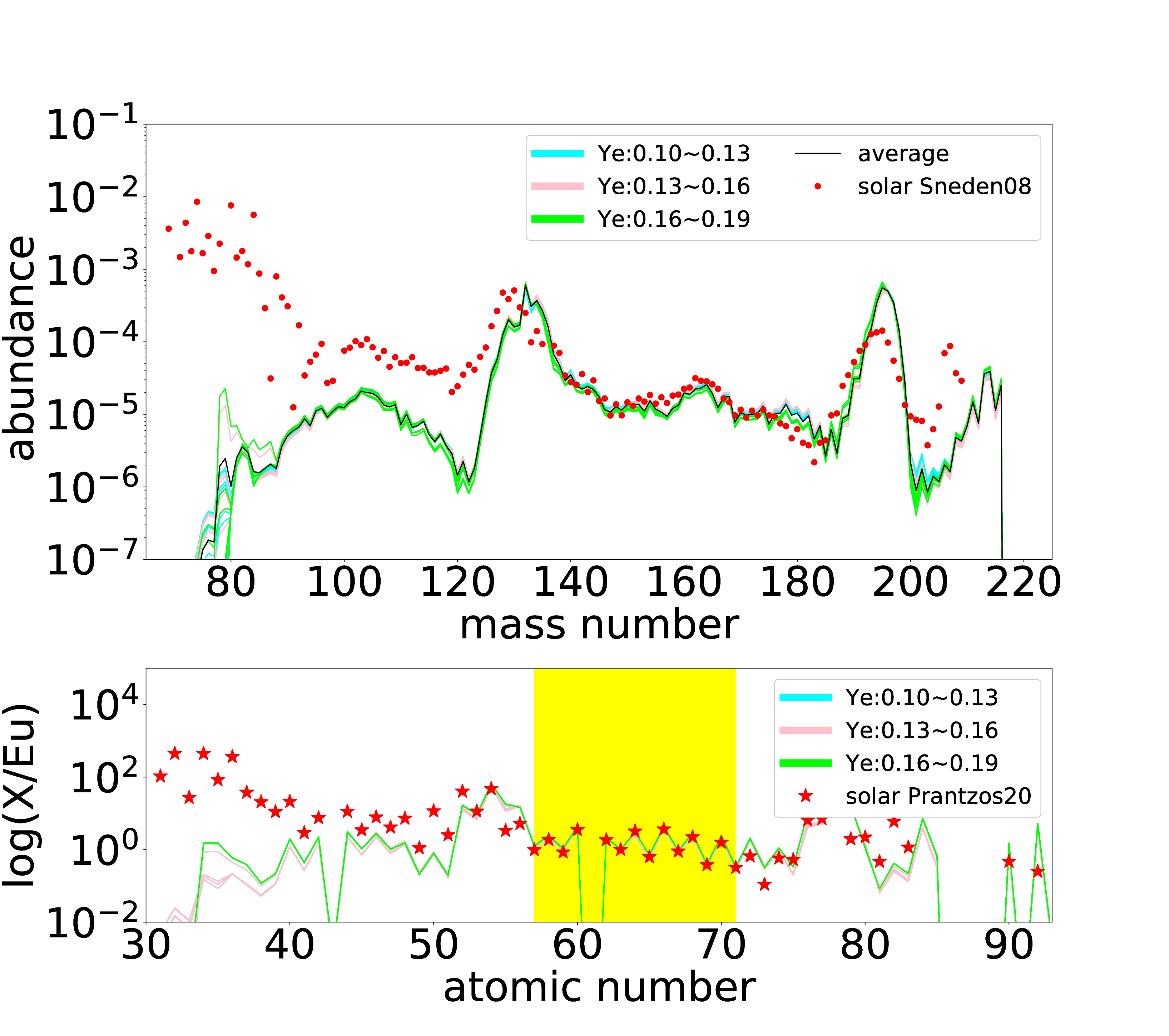}
\caption{Top panel: the selected abundance pattern of the calculated CEJSN r-process nucleosynthesis grouped by the initial electron fraction: $0.1 \le Y_{\rm e} < 0.13$ in cyan, i.e., the cyan lines show different values of $Y_{\rm e}$ in that range, $0.13 \le Y_{\rm e}<0.16$ (pink), and $0.16 \le Y_{\rm e}<0.19$ (green) as function of mass number. The black solid line is the average of these groups. The observed solar system abundances as a function of mass number are shown with red dots(\cite{sneden08}). Bottom panel: the abundance as a function of atomic number of the selected groups as in the top panel and the observed solar abundance(\cite{prantzos19}). The yellow area represents the region of universality of the r-process (see text). }
\label{abund}
\end{figure}

The universality of this work is shown in the yellow region of the bottom panel in Fig \ref{abund}.
We select 21 abundances pattern as a function of the atomic number and classify them into three groups by the initial electron fraction,  
$0.1 \le Y_{\rm e} < 0.13$,  $0.13 \le Y_{\rm e}<0.16$ , and $0.16 \le Y_{\rm e}<0.19$.
All these selected patterns favor the fastest expansion and highest entropy of the input parameters, which eventually lead the nucleosynthesis flow to heavy regions. 
As the criterion of the selection, elements of atomic number 56 $\le Z \le$ 70 match the solar abundance from the paper (\cite{sneden08}) pretty well.   
Elements of $71 \le Z \lesssim 80$ from our nucleosynthesis partially match the solar abundance, while our results are no longer consistent with the solar abundance for $80 \le Z$.
More detailed investigation of this remarkable feature of the CEJSN will be discussed with the observation of r-enhanced metal-poor star in section \ref{sec:Xla}.  

A recent research found that the universality also appears in the light elements region(\cite{Roederer_2022}), it is not observed in the present work because the CEJSN essentially favor the elemental production in the area beyond the second peak. 
The abundances from the present work as a function of mass number are classified in three groups with varying $Y_{\rm e}$ and their average abundance are compared with the solar abundance(\cite{sneden08}). 
Besides perfect matching with the solar data on the Lanthanide region of 140$\le A \le$175, a robust elemental pattern in the second and third peaks of the r-process is also produced. This indicates that the CEJSN provides a promising site of the strong r-process, also referred as main r-process.
The strength of the second peak from the CEJSN scenario is comparable with the solar abundance, but it is slightly displaced towards the right a few mass units. 
The second peak can be produced both in the weak r-process and in the strong r-process with different synthesizing mechanisms. 
In the weak r-process the ($\alpha$,n) reactions take over the dominance from the neutron capture reactions(\cite{bliss2017,Jin_2022}) to form the second peak. 
On the other hand, in the strong r-process scenario, the r-process can reach to extremely heavy nuclei of the neutron-shell closure N = 184, where the fission fragments are formed in the mass number range from 115 to 155 (\cite{Eichler_2015}) and hence cover the area of the second peak. 
In this work, the fission process is not observed during the nucleosynthesis since no oscillating over the average atomic number as function of evolution time is found, which means no fission circles happened. 
This is consistent with previous research because the Ye of current work is larger than 0.1, while the fission process usually happens at extremely low Ye around 0.05(\cite{Korobkin_2012}).

\subsection{Lanthanides and the Third peak}
\label{sec:Xla}

\begin{figure}
\centering
\includegraphics[trim=0cm 0 0 0, clip, width=0.48\textwidth]{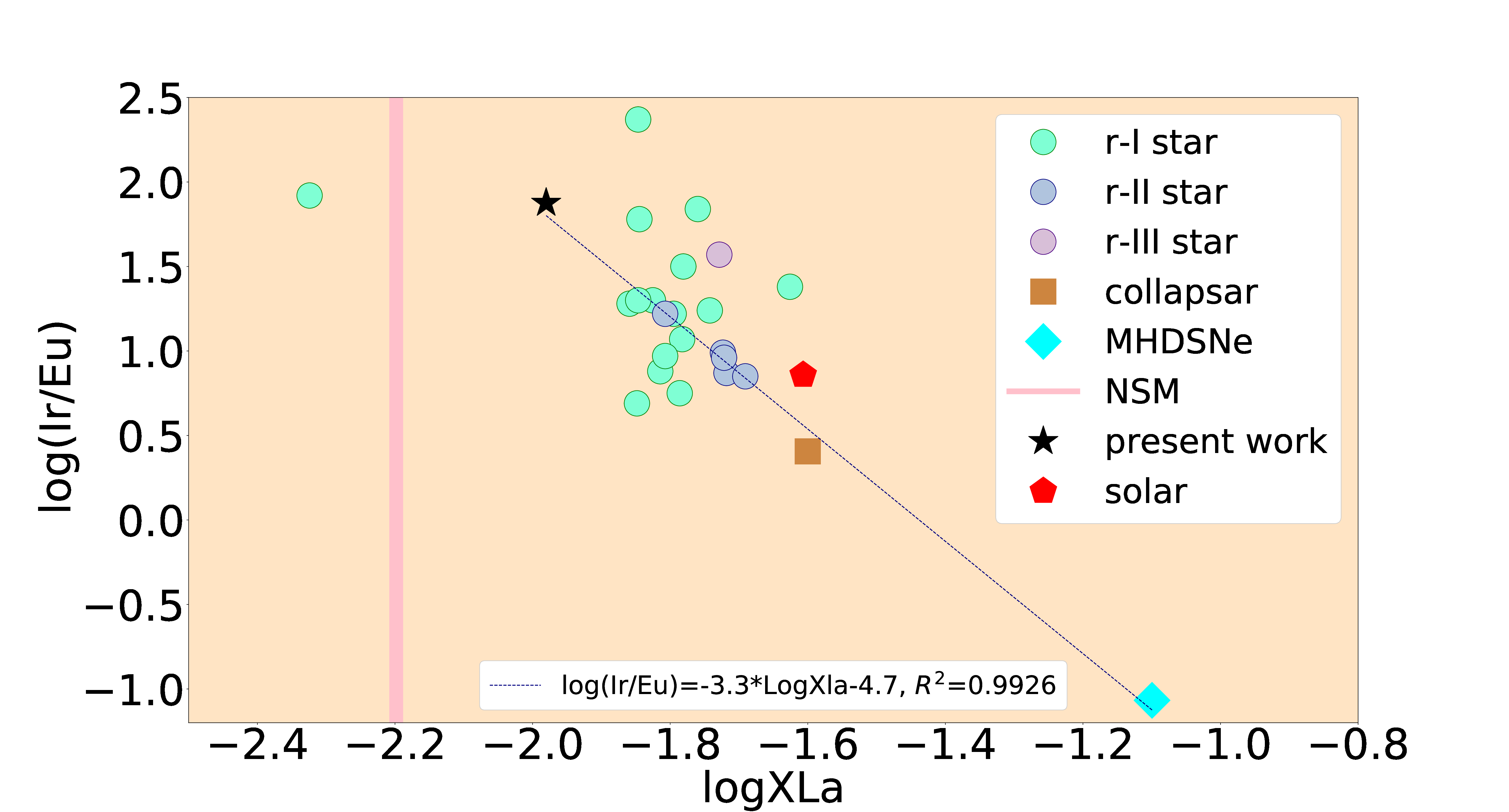}
\caption{$\log{\rm(Ir/Eu)}$ versus $\log{\rm XLa}$ of r-enhanced stars, the solar system, and r-process theoretical calculations of the collapsar scenario, the MHDSN scenario, and the present CEJSN r-process scenario. XLa is the ratio between the Lanthanides mass and the total mass of r-process elements. The data of the r-enhanced stars are extracted from JINAbase(\cite{Abohalima_2018}) with the two constraints of [Fe/H] $<$ -2.5 and [Ba/Eu] $<$ -0.4.  We present the mean value of the $\dot{M}_{1}$ and $\dot{M}_{2}$ accretion regimes  from the collapsar model(\cite{collapsar}), which represent the strong r-process components. The mean value of the top three strongest r-process trace of the MHDSN scenario(\cite{Reichert_mnras}) are presented. The dashed line shows the liner fitting of the MHD, the collapsar and the CEJSN scenarios with ${R^2}$=0.9926. }
\label{xlaIrEu}
\end{figure}

The most noteworthy feature to notice of the CEJSN r-process nucleosynthesis is the abundance patterns in the region beyond the Lanthanides: the calculated third peak is about ten times stronger than the solar abundance while the Lanthanides are similar to the solar abundance.
To demonstrate the strength of the enhanced third peak more precisely we choose one representative element from the Lanthanide and another element from the third peak and compare their relative abundance.
The Iridium, Z=77, is chosen to represent the third peak since it is the most observed element in stars.
For the Lanthanide, the natural element is Europium (Z=63) because it is a nearly pure r-process element and most readily measurable in optical spectra.
The results of the CEJSN r-process scenario are presented in Fig.\ref{xlaIrEu}, alongside theoretical results from other r-process models and observations of metal-poor stars in the plane of log(Ir/Eu) versus log(XLa), where XLa is the ratio between the Lanthanides mass and the total mass of r-process elements. 

The metal-poor stars that were born in the early Universe were enriched by one to a few r-process nucleosynthesis events. They have maintained their abundance signatures over billions of years without pollution, and therefore provide significant fingerprints to constrain the theoretical r-process scenarios.
In this work, the chemical abundances of Ir and Eu are extracted from the JINAbase(\cite{Abohalima_2018}), which is a large database along with a queryable web application of 1659 metal poor stars. 
Two more conditions are added in selecting the target stars, ${\rm [Fe/H]} < -2.5$ to have metal-poor stars and ${\rm [Ba/Eu]} < -0.4$ to have pure r-process stars.
Finally, 16 r-I stars, 5 r-II stars and one r-III stars are picked out. 

As can be seen in the Fig \ref{xlaIrEu}, there is a broad range from log(Ir/Eu)=0.69 to log(Ir/Eu)=2.37 of the r-I, a much smaller distribution from 0.85 to 1.22 of the r-II stars, and the unique r-III star has log(Ir/Eu)=1.57. Overall, $86 \%$ of these r-enhanced stars have higher log(Ir/Eu) than the value in the solar system(\cite{sneden08}).  
The mean value of the $\dot{M}_{1}$ and $\dot{M}_{2}$ accretion regimes of the collapsar scenario(\cite{collapsar}), which yield the strong r-process in that scenario, is $\log{\rm (Ir/Eu)}=0.41$(\cite{Ji_2019}). 
The mean value of the top three strong r-process abundance patterns of the MHDSNe scenario from (\cite{oberg/mnrasl/slx046,Reichert_mnras}) is $\log{\rm (Ir/Eu)}=-1.07$, which is the only negative value in Fig \ref{xlaIrEu}.  
The log(Ir/Eu) of the present CEJSN r-process scenario calculation is 1.87. It is the highest value among the theoretical models, while only two r-I metal poor stars have larger values.
This high Ir/Eu ratio clearly demonstrates that the CEJSN scenario might be the unique model for explaining the abundant elements in the third peak of the observed r-enhanced stars

Fig.\ref{xlaIrEu} is informative in presenting logXLa that represents the  Lanthanides mass in addition to log(Ir/Eu) that represents the relative strength of the third r-process peak. 
The value for the kilonova of the neutron star merger event GW170817 is $\log{\rm XLa} = -2.2$(\cite{Ji_2019}), as the pink line indicates in Fig. \ref{xlaIrEu}. 
It is several times smaller than the statistical mean of $(\log{\rm XLa})_{\rm r-star} = -1.55$ of 146 r-enhanced stars(\cite{Ji_2019}). That means that more lanthanide rich, i.e., $\log{\rm XLa}> -1.5$, NSM events should be observed in coming years if the NSM is a dominant r-process site. 
It seems that the NSM scenario can not supply enough lanthanides based on the only observed NSM event GW170817 so far. 
However, other r-process models can. The MHDSNe scenario gives the largest value of $\log{\rm XLa}= -1.09$, while the value of the collapsar scenario is  $\log{\rm XLa} \simeq -1.6$, very close to the solar system value. 
In the CEJSN scenario $\log{\rm XLa} = -1.98$, slightly larger than the value of the GW170817 kilonova.
These three r-process sites, the MHDSNe scenario, the collapsar scenario and the CEJSN scenario, are basically on a straight line with a slope of -3.3 in Fig.\ref{xlaIrEu}.
Although the samples of r-enhanced metal-poor stars do not share the same slope, they do have a general similar tendency of an increase in $\log{\rm (Ir/Eu)}$ as $\log{\rm XLa}$ decreases. 
Comparing each of the r-process models with the r-enhanced stars in the fitting line in Fig. \ref{xlaIrEu}, we thus infer that the production of large quantities of Lanthanides and of the third r-process peak can not occur in a single strong r-process site.

In the right end of the fitting line, models like MHDSNe can produce the lanthanides that account for more the 10\% of the r-process mass, but this scenario barely contributes to elements in the third peak. 
Most metal-poor stars which don't have the abundance of elements in the third peak are located near the collapsar in Fig.\ref{xlaIrEu}. For example, the  Ultra-faint dwarf galaxy TucanaIII stars have $\log {\rm XLa} \simeq -1.5$ but no elements observed in the third peak(\cite{Marshall_2019}). However, many r-enhanced stars have higher $\log({\rm Ir/Eu})$ and lower $\log {\rm XLa}$ values than what the collapsar scenario gives, and closer to what the present work shows for the CEJSN r-process scenario.  
The universality of Lanthanides was used by previous researchers  to identify r-process nucleosynthesis in stars. They emphasize the relative abundances as a proof of r-process.
Since the ratio of Lanthanides to total r-process elements varies by about an order of magnitude between different r-process models, it is crucial to use $\log {\rm XLa}$ combined with the third peak of the r-process to determine the r-process models. The anti-correlation shown in the fitting line in Fig.\ref{xlaIrEu} reveals an essential feature of the r-process, namely, that it can be a significant constraint for further study on the nucleosynthesis from Lanthanides to the third peak . 

The CEJSN is extremely important in this conclusion because it is the only model with high $\log{\rm (Ir/Eu)}$ and low $\log{\rm XLa}$, thus it provides a reasonable explanation to some r-enhanced stars which other r-process scenarios cannot access.
It covers the observed r-process-enhanced stars in sub-classes r-II and r-III and quite a number of r-I stars. Even for those extreme cases, like the most Lanthanides poor star CS30312-044(\cite{Roederer_2014}) (the most-left star in Fig. \ref{xlaIrEu}) CEJSN produce almost the same mass of the third r-process peak, and thus explain this particular star at least in this respect. The star CS29513-003(\cite{Roederer_2014}) that has the largest component of the third peak among the r-enhanced stars (upper-most star in Fig. \ref{xlaIrEu}) and a relative mild value of $\log{\rm (XLa)}$ is compatible with the trend that strong third peak stars are anti-correlated with high production of Lanthanides. It is not far from the prediction of the present CEJSN r-process scenario.
The reason the CEJSN r-process scenario accounts for high $\log({\rm Ir/Eu})$ is the density function of CEJSN jets according to equation (\ref{eq:DensityEX2}) that has large density of neutrons in the late phase of the r-process nucleosynthesis, implying production of isotopes with larger mass numbers more efficiently.

\subsection{Cosmochronometry}
\label{sec:cosm}
 The CEJSNe can also be very promising in accounting for Actinide-boosted stars for same reason. The star 2MASS J09544277+5246414 (\cite{Holmbeck_2018}), which is the most Actinide-enhanced r-II star observed so far with [Th/Eu] = +0.37, has unattainable estimated age from other r-process models with cosmochronometry. The equation of cosmochronometry(\cite{Schatz_2002}) reads 
\begin{equation}  
\tau = 46.7 \times [\log({\rm Th/Eu})_{0} - \log({\rm Th/Eu})_{*} ]~ {\rm Gyr}
\label{eq:cosmochronometr}
\end{equation}
where $\log({\rm Th/Eu})_{0}$ is calculated from the theoretical model and $\log({\rm Th/Eu})_{*}$ is the observed value in the star.
The $\log({\rm Th/Eu})_{0}$ from the present CEJSN model is 0.12, which gives an age of 11.09 Gyr for this star, compatible with the age of the Universe. 
Generally, the CEJSN scenario supplies multiple perspectives to constrain and understand the origin of heavy elements, especially above the Lanthanides, in the early Universe.

\section{Conclusion}
\label{sec:concl}

We initially report the elemental abundance pattern of r-process in the CEJSN scenario. Due to the high production of elements beyond the Lanthanides, the CEJSN could be the best candidate for explaining the characteristic of r-enhanced metal-poor stars among all r-process sites.
We also propose a new quantity, Log(Xla) VS Log(Ir/Eu), which reflects the relative production of the third peak elements to the Lanthanides in r-process. We find anti-correlation of high fraction of Lanthanides to the third peak of current r-process models. This would be a critical feature for further study on r-process.
Since considerable production of the Actinides, the CEJSN effectively explains the age the most Actinide-boost star by the cosmochronometry.
The uncertainties of the nuclear input, like mass, reaction rate and decay properties will be studied in future work.

\acknowledgments
Shilun Jin is supported by Major State Basic Research Development Program of China No.2018YFA0404401, National Natural Science Foundation of China No.12375145, CAS-Light of West China Program No.2020-82 and CAS Project for Young Scientists in Basic Research No.YSBR-002.
Noam Soker is supported by a grant from the Israel Science Foundation (769/20).


\begin{thebibliography}{}
\expandafter\ifx\csname natexlab\endcsname\relax\def\natexlab#1{#1}\fi
\providecommand{\url}[1]{\href{#1}{#1}}
\providecommand{\dodoi}[1]{doi:~\href{http://doi.org/#1}{\nolinkurl{#1}}}
\providecommand{\doeprint}[1]{\href{http://ascl.net/#1}{\nolinkurl{http://ascl.net/#1}}}
\providecommand{\doarXiv}[1]{\href{https://arxiv.org/abs/#1}{\nolinkurl{https://arxiv.org/abs/#1}}}

\bibitem[{Abbott {et~al.}(2017)Abbott, Abbott, \& Abbott.{\it et al.}}]{Abbott_2017}
Abbott, B.~P., Abbott, R., \& Abbott.{\it et al.}, T.~D. 2017, Astrophys. J., 848, L12, \dodoi{10.3847/2041-8213/aa91c9}

\bibitem[{Abbott.{\it et al.}(2017)}]{PhysRevLett-119-161101}
Abbott.{\it et al.}, B.~P. 2017, Phys. Rev. Lett., 119, 161101, \dodoi{10.1103/PhysRevLett.119.161101}

\bibitem[{Abohalima \& Frebel(2018)}]{Abohalima_2018}
Abohalima, A., \& Frebel, A. 2018, The Astrophy. J. S., 238, 36, \dodoi{10.3847/1538-4365/aadfe9}

\bibitem[{Akashi \& Soker(2021)}]{Akashi_2021}
Akashi, M., \& Soker, N. 2021, Astrophy.J., 923, 55, \dodoi{10.3847/1538-4357/ac2d2b}

\bibitem[{Arcones \& Thielemann(2012)}]{Arcones_2013}
Arcones, A., \& Thielemann, F.-K. 2012, J. Phys. G, 40, 013201, \dodoi{10.1088/0954-3899/40/1/013201}

\bibitem[{Beers \& Christlieb(2005)}]{metalpoor}
Beers, T.~C., \& Christlieb, N. 2005, Ann. Rev. Astrono. Astrophy., 43, 531, \dodoi{10.1146/annurev.astro.42.053102.134057}

\bibitem[{{Beniamini} \& {Piran}(2019)}]{BeniaminiPiran2019}
{Beniamini}, P., \& {Piran}, T. 2019, Mon. Not. R. Astro. Soc., 487, 4847, \dodoi{10.1093/mnras/stz1589}

\bibitem[{Bliss {et~al.}(2017)Bliss, Arcones, Montes, \& Pereira}]{bliss2017}
Bliss, J., Arcones, A., Montes, F., \& Pereira, J. 2017, J. Phys. G, 44, 054003, \dodoi{10.1088/1361-6471/aa63bd}

\bibitem[{Burbidge {et~al.}(1957)Burbidge, Burbidge, Fowler, \& Hoyle}]{B2FH}
Burbidge, E.~M., Burbidge, G.~R., Fowler, W.~A., \& Hoyle, F. 1957, Rev. Mod. Phys., 29, 547, \dodoi{10.1103/RevModPhys.29.547}

\bibitem[{Burrows \& Vartanyan(2021)}]{Burrow21}
Burrows, A., \& Vartanyan, D. 2021, Nature, 589, 29, \dodoi{10.1038/s41586-020-03059-w}

\bibitem[{Chornock {et~al.}(2017)Chornock, Berger, \& Kasen.{\it et al.}}]{Chornock_2017}
Chornock, R., Berger, E., \& Kasen.{\it et al.}, D. 2017, Astrophys. J., 848, L19, \dodoi{10.3847/2041-8213/aa905c}

\bibitem[{Coulter {et~al.}(2017)Coulter, Foley, \& Kilpatrick.{\it et al.}}]{coulter17}
Coulter, D.~A., Foley, R.~J., \& Kilpatrick.{\it et al.}, C.~D. 2017, Science, 358, 1556, \dodoi{10.1126/science.aap9811}

\bibitem[{Cyburt {et~al.}(2010)Cyburt, Amthor, \& Ferguson.{\it et al.}}]{Cyburt_2010}
Cyburt, R.~H., Amthor, A.~M., \& Ferguson.{\it et al.}, R. 2010, Astrophy. J., 189, 240, \dodoi{10.1088/0067-0049/189/1/240}

\bibitem[{Côté {et~al.}(2019)Côté, Eichler, \& Arcones.{\it et al.}}]{Côté_2019}
Côté, B., Eichler, M., \& Arcones.{\it et al.}, A. 2019, Astrophy. J, 875, 106, \dodoi{10.3847/1538-4357/ab10db}

\bibitem[{Ebinger {et~al.}(2018)Ebinger, Curtis, \& Fröhlich.{\it et al.}}]{Ebinger_2019}
Ebinger, K., Curtis, S., \& Fröhlich.{\it et al.}, C. 2018, Astrophys. J, 870, 1, \dodoi{10.3847/1538-4357/aae7c9}

\bibitem[{Eichler {et~al.}(2015)Eichler, Arcones, \& Kelic.{\it et al.}}]{Eichler_2015}
Eichler, M., Arcones, A., \& Kelic.{\it et al.}, A. 2015, Astrophys. J., 808, 30, \dodoi{10.1088/0004-637X/808/1/30}

\bibitem[{Frebel(2018)}]{Frebel18}
Frebel, A. 2018, Annual Review of Nuclear and Particle Science, 68, 237, \dodoi{10.1146/annurev-nucl-101917-021141}

\bibitem[{Grichener {et~al.}(2022)Grichener, Kobayashi, \& Soker}]{Grichener_2022}
Grichener, A., Kobayashi, C., \& Soker, N. 2022, Astrophy. J., 926, L9, \dodoi{10.3847/2041-8213/ac4f68}

\bibitem[{Grichener \& Soker(2019{\natexlab{a}})}]{Grichener_2019}
Grichener, A., \& Soker, N. 2019{\natexlab{a}}, Astrophy. J, 878, 24, \dodoi{10.3847/1538-4357/ab1d5d}

\bibitem[{Grichener \& Soker(2019{\natexlab{b}})}]{grichener2019paradigm}
---. 2019{\natexlab{b}}.
\newblock \doarXiv{1909.06328}

\bibitem[{Grichener \& Soker(2021)}]{10.1093/mnras/stab2233}
---. 2021, Mon. Not. R. Astro. Soc., 507, 1651, \dodoi{10.1093/mnras/stab2233}

\bibitem[{Holmbeck \& Andrews(2023)}]{holmbeck_2023}
Holmbeck, E.~M., \& Andrews, J.~J. 2023, Total r-process Yields of Milky Way Neutron Star Mergers.
\newblock \doarXiv{2310.03847}

\bibitem[{Holmbeck {et~al.}(2018)Holmbeck, Beers, \& Roederer.{\it et al.}}]{Holmbeck_2018}
Holmbeck, E.~M., Beers, T.~C., \& Roederer.{\it et al.}, I.~U. 2018, Astrophy. J., 859, L24, \dodoi{10.3847/2041-8213/aac722}

\bibitem[{Holmbeck {et~al.}(2020)Holmbeck, Hansen, \& Beers.{\it et al.}}]{Holmbeck_2020}
Holmbeck, E.~M., Hansen, T.~T., \& Beers.{\it et al.}, T.~C. 2020, Astrophy. J. S., 249, 30, \dodoi{10.3847/1538-4365/ab9c19}

\bibitem[{Ji {et~al.}(2019)Ji, Drout, \& Hansen}]{Ji_2019}
Ji, A.~P., Drout, M.~R., \& Hansen, T.~T. 2019, Astrophy. J, 882, 40, \dodoi{10.3847/1538-4357/ab3291}

\bibitem[{Jin(2022)}]{Jin_2022}
Jin, S. 2022, Astrophy. J., 927, 116, \dodoi{10.3847/1538-4357/ac4f4a}

\bibitem[{Kasen {et~al.}(2017)Kasen, Metzger, \& Barnes.{\it et al.}}]{kasen17}
Kasen, D., Metzger, B., \& Barnes.{\it et al.}, J. 2017, Nature, 551, 80, \dodoi{10.1038/nature24453}

\bibitem[{Kasliwal {et~al.}(2019)Kasliwal, Kasen, \& Lau.{\it et al.}}]{kasliwal19}
Kasliwal, M.~M., Kasen, D., \& Lau.{\it et al.}, R.~M. 2019, Mon. Not. R. Astro. Soc., 510, L7, \dodoi{10.1093/mnrasl/slz007}

\bibitem[{{Kobayashi} {et~al.}(2020){Kobayashi}, {Karakas}, \& {Lugaro}}]{Kobayashietal2020}
{Kobayashi}, C., {Karakas}, A.~I., \& {Lugaro}, M. 2020, \apj, 900, 179, \dodoi{10.3847/1538-4357/abae65}

\bibitem[{{Kobayashi} {et~al.}(2023){Kobayashi}, {Mandel}, {Belczynski}, {Goriely}, {Janka}, {Just}, {Ruiter}, {Vanbeveren}, {Kruckow}, {Briel}, {Eldridge}, \& {Stanway}}]{Kobayashietal2023}
{Kobayashi}, C., {Mandel}, I., {Belczynski}, K., {et~al.} 2023, \apj, 943, L12, \dodoi{10.3847/2041-8213/acad82}

\bibitem[{Korobkin {et~al.}(2012)Korobkin, Rosswog, Arcones, \& Winteler}]{Korobkin_2012}
Korobkin, O., Rosswog, S., Arcones, A., \& Winteler, C. 2012, Monthly Notices of the Royal Astronomical Society, 426, 1940, \dodoi{10.1111/j.1365-2966.2012.21859.x}

\bibitem[{{Lippuner} \& {Roberts}(2017)}]{skynet_2017}
{Lippuner}, J., \& {Roberts}, L. 2017, Astrophy. J. S., 233, 18, \dodoi{https://doi.org/10.3847/1538-4365/aa94cb}

\bibitem[{Marshall.{\it et al.}(2019)}]{Marshall_2019}
Marshall.{\it et al.}, J.~L. 2019, The Astrophy. J., 882, 177, \dodoi{10.3847/1538-4357/ab3653}

\bibitem[{Montes {et~al.}(2007)Montes, Beers, \& Cowan.{\it et al.}}]{Montes2007}
Montes, F., Beers, T.~C., \& Cowan.{\it et al.}, J. 2007, Astrophys. J, 671, 1685, \dodoi{10.1086/523084}

\bibitem[{Mumpower {et~al.}(2016)Mumpower, McLaughlin, Surman, \& Steiner}]{Mumpower_2016}
Mumpower, M.~R., McLaughlin, G.~C., Surman, R., \& Steiner, A.~W. 2016, Astrophy. J., 833, 282, \dodoi{10.3847/1538-4357/833/2/282}

\bibitem[{Mösta {et~al.}(2018)Mösta, Roberts, \& Halevi.{\it et al.}}]{mosta2018}
Mösta, P., Roberts, L.~F., \& Halevi.{\it et al.}, G. 2018, Astrophys. J, 864, 171, \dodoi{10.3847/1538-4357/aad6ec}

\bibitem[{Obergaulinger \& Aloy(2021)}]{oberg2021_3D}
Obergaulinger, M., \& Aloy, M. 2021, Mon. Not. R. Astro. Soc., 503, 4942, \dodoi{10.1093/mnras/stab295}

\bibitem[{Obergaulinger \& Aloy(2017)}]{oberg/mnrasl/slx046}
Obergaulinger, M., \& Aloy, M.~A. 2017, Mon. Not. R. Astro. Soc., 469, L43, \dodoi{10.1093/mnrasl/slx046}

\bibitem[{{Papish} {et~al.}(2015){Papish}, {Soker}, \& {Bukay}}]{Papishetal2015}
{Papish}, O., {Soker}, N., \& {Bukay}, I. 2015, Mon. Not. R. Astro. Soc., 449, 288, \dodoi{10.1093/mnras/stv345}

\bibitem[{Pian {et~al.}(2017)Pian, D’Avanzo, \& Benetti.{\it et al.}}]{pian17}
Pian, E., D’Avanzo, P., \& Benetti.{\it et al.}, S. 2017, Nature, 551, 67, \dodoi{10.1038/nature24298}

\bibitem[{Prantzos {et~al.}(2019)Prantzos, Abia, \& Cristallo.{\it et al.}}]{prantzos19}
Prantzos, N., Abia, C., \& Cristallo.{\it et al.} 2019, Mon. Not. R. Astro. Soc., 491, 1832, \dodoi{10.1093/mnras/stz3154}

\bibitem[{Radice {et~al.}(2018)Radice, Perego, \& Hotokezaka.{\it et al.}}]{Radice_2018}
Radice, D., Perego, A., \& Hotokezaka.{\it et al.}, K. 2018, Astrophy. J., 869, 130, \dodoi{10.3847/1538-4357/aaf054}

\bibitem[{Reichert {et~al.}(2021)Reichert, Obergaulinger, Eichler, Aloy, \& Arcones}]{Reichert_mnras}
Reichert, M., Obergaulinger, M., Eichler, M., Aloy, M.~A., \& Arcones, A. 2021, Mon. Not. R. Astro. Soc., 501, 5733, \dodoi{10.1093/mnras/stab029}

\bibitem[{Roederer {et~al.}(2022)Roederer, Cowan, \& Pignatari.{\it et al.}}]{Roederer_2022}
Roederer, I.~U., Cowan, J.~J., \& Pignatari.{\it et al.}, M. 2022, Astrophy. J., 936, 84, \dodoi{10.3847/1538-4357/ac85bc}

\bibitem[{Roederer {et~al.}(2014)Roederer, Preston, \& Thompson.{\it et al.}}]{Roederer_2014}
Roederer, I.~U., Preston, G.~W., \& Thompson.{\it et al.}, I.~B. 2014, Astrono. J., 147, 136, \dodoi{10.1088/0004-6256/147/6/136}

\bibitem[{Schatz {et~al.}(2002)Schatz, Toenjes, \& Pfeiffer.{\it et al.}}]{Schatz_2002}
Schatz, H., Toenjes, R., \& Pfeiffer.{\it et al.}, B. 2002, Astrophy. J., 579, 626, \dodoi{10.1086/342939}

\bibitem[{Siegel {et~al.}(2019)Siegel, Barnes, \& Metzger}]{collapsar}
Siegel, D., Barnes, J., \& Metzger, B. 2019, Nature, 569, 241, \dodoi{10.1038/s41586-019-1136-0}

\bibitem[{Siegel(2022)}]{Siegel22}
Siegel, D.~M. 2022, Nature Reviews Physics, 4, 306, \dodoi{doi.org/10.1038/s42254-022-00439-1}

\bibitem[{Sneden {et~al.}(2008)Sneden, Cowan, \& Gallino}]{sneden08}
Sneden, C., Cowan, J.~J., \& Gallino, R. 2008, Ann. Rev. Astrono. Astrophy., 46, 241, \dodoi{10.1146/annurev.astro.46.060407.145207}

\bibitem[{{Tarumi} {et~al.}(2021){Tarumi}, {Hotokezaka}, \& {Beniamini}}]{Tarumietal2021}
{Tarumi}, Y., {Hotokezaka}, K., \& {Beniamini}, P. 2021, \apj, 913, L30, \dodoi{10.3847/2041-8213/abfe13}

\bibitem[{Truran {et~al.}(2002)Truran, Cowan, Pilachowski, \& Sneden}]{Truran2002}
Truran, J.~W., Cowan, J.~J., Pilachowski, C.~A., \& Sneden, C. 2002, Pub. Astro. Soc. Pac., 114, 1293, \dodoi{10.1086/344585}

\bibitem[{Villar {et~al.}(2017)Villar, Guillochon, \& Berger.{\it et al.}}]{Villar_2017}
Villar, V.~A., Guillochon, J., \& Berger.{\it et al.}, E. 2017, Astrophys. J., 851, L21, \dodoi{10.3847/2041-8213/aa9c84}

\bibitem[{Watson {et~al.}(2019)Watson, Hansen, \& Selsing.{\it et al.}}]{Sr19}
Watson, D., Hansen, C.~J., \& Selsing.{\it et al.}, J. 2019, Nature, 574, 497, \dodoi{doi.org/10.1038/s41586-019-1676-3}

\bibitem[{{Waxman} {et~al.}(2018){Waxman}, {Ofek}, {Kushnir}, \& {Gal-Yam}}]{Waxmanetal2018}
{Waxman}, E., {Ofek}, E.~O., {Kushnir}, D., \& {Gal-Yam}, A. 2018, Mon. Not. R. Astro. Soc., 481, 3423, \dodoi{10.1093/mnras/sty2441}

\bibitem[{Yong {et~al.}(2021)Yong, Kobayashi, \& Costa.{\it et al.}}]{Yong2021}
Yong, D., Kobayashi, C., \& Costa.{\it et al.}, G. S.~D. 2021, Nature, 595, 223, \dodoi{10.1038/s41586-021-03611-2}

\end{thebibliography}
\end{document}